\documentclass[preprint,12pt]{elsarticle}
\usepackage[utf8]{inputenc}
\usepackage{orcidlink}
\usepackage[T1]{fontenc}
\usepackage{amsmath}
\usepackage{hyperref}

\title{Nonreciprocal topological kink-wave propagation in mechanical metamaterials} 
\author{Brahim Lemkalli$^{1}$, Qingxiang Ji$^{1,\ast}$, Jingyi Zhang$^{2}$, Richard Craster$^{3}$, Johan Christensen$^{2}$ and Muamer Kadic$^{1}$\\
\small$^{1}$Université Marie et Louis Pasteur, SUPMICROTECH, Institut FEMTO-ST, 25000 Besançon, France\\
\small$^{2}$IMDEA Materials Institute, 28906, Getafe, Madrid, Spain\\
\small$^{3}$UMI 2004 Abraham de Moivre-CNRS, Imperial College London, SW7~2AZ, London, UK \\
\small$^\ast$Corresponding author. Email: qingxiang.ji@femto-st.fr}

\date{}
\begin{document}

\begin{abstract}
Nonlinear mechanical metamaterials can exhibit emergent transport phenomena that mimic topological protection without relying on linear band topology. Here, we realize a bifurcation-induced nonreciprocal lattice that supports robust propagation of elastic kink waves. Each unit is a prestrained, hinged-beam circulator that develops angular-momentum bias during snap-through transitions between buckling states, producing an effective breaking of time-reversal symmetry. Coupling such units into a hexagonal array yields a mechanically chiral network where localized soliton-like excitations propagate unidirectionally along interfaces and edges, immune to sharp bends. We demonstrate non-dispersive kink transport governed by a Sine–Gordon-type field whose effective bias encodes mechanical chirality. This framework bridges bifurcation dynamics and nonreciprocal transport, establishing a nonlinear route toward topological-like mechanical functionality without magnetic or gyroscopic bias.
\end{abstract}
\maketitle 
The concept of topological order, first discovered in condensed matter physics, has inspired the study of topologically nontrivial states across various physical systems \cite{konig2007quantum,moore2010birth,bernevig2006quantum,huber2016topological,wen2019choreographed,coulais2021topology}. These states underlie phenomena such as the quantum Hall effect, topological insulators \cite{haldane1988model,hasan2010colloquium,susstrunk2015observation,zheng2024three,chen2019topological,chaplain2020topological,song2020twisted}, and their photonic analogues, including photonic crystals and chiral waveguides \cite{li2020higher,zangeneh2019nonlinear,qin2025disorder,chen2023observation,song2019topological}. Their robustness against defects enables novel wave propagation functionalities \cite{chen2018elastic,zhang2023second}. Crucially, edge states traverse sharp bends without backscattering, advancing topological wave guiding \cite{sessi2016robust,fleury2016floquet,ghatak2020observation,zhang2019deep,chaunsali2021stability}.  
Topological protection in condensed matter often relies on time-reversal symmetry or non-symmetry \cite{qi2011topological,jin2017infrared,zhang2019non}. Breaking this symmetry enables nonreciprocal edge states in quantum Hall and photonic systems \cite{sounas2017non,khanikaev2015topologically,jin2016topological}, while bosonic systems require alternative mechanisms, as acoustic waves lack strong magnetic interactions \cite{peri2020experimental}. We extend these principles to mechanical metamaterials using angular momentum bias to break time-reversal symmetry, achieving protected, nonreciprocal elastic states in pre-strained systems \cite{amendola2018tuning,dudek2025shape}.

\begin{figure}
    \centering    \includegraphics[width=0.9\linewidth]{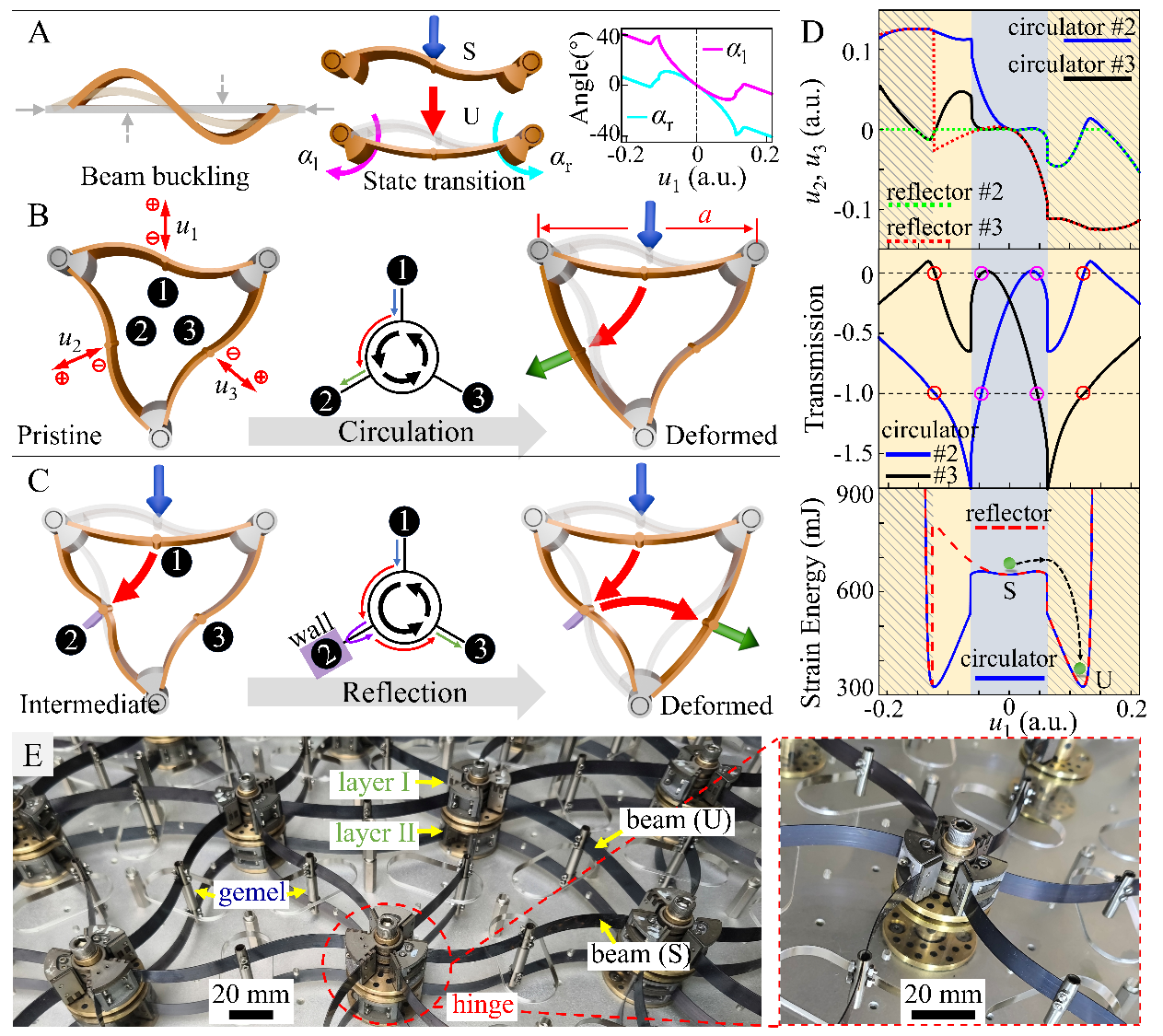}
    \caption{\textbf{Mechanical circulator and reflector.} (A) Straight flexible beam with pinned ends in secondary buckling (S-mode) snaps through to primary buckling (U-mode) under midpoint load (blue arrow). During snap-through, left hinge rotation $\alpha_l$ (magenta) exceeds right hinge rotation $\alpha_r$ (cyan), as shown versus normalized displacement ($u_1/a$) in the right panel. Beam: length $105~\mathrm{mm}$, width $10~\mathrm{mm}$, thickness $0.5~\mathrm{mm}$, hinge radius $20~\mathrm{mm}$, pre-compressed by $5~\mathrm{mm}$ to span $140~\mathrm{mm}$. (B) Mechanical circulator formed by three hinged beams in a triangular loop. Compression towards the centroid induces secondary buckling, producing a chiral three-port system. A load at Port 1 triggers $1\!\to\!2$ circulation (Port 2 active, Port 3 passive) via momentum biasing; similarly for $2\!\to\!3$ and $3\!\to\!1$. (C) Reflector created by adding a hard wall (purple) at Port 2, restricting motion inward. A load at Port 1 redirects displacement to Port 3 (mechanical reflection), while a load at Port 3 outputs directly to Port 1. (D) Port displacements ($u_1/a$): circulator—Port 2 (blue), Port 3 (black); reflector—Port 2 (green dashed), Port 3 (red dashed). Blue/yellow regions mark circulator pre/post-snap; unshaded/shaded mark reflector pre/post-snap. For $u_1>0$, responses coincide; for $u_1<0$, reflector snap-through is delayed. Middle: transmission before (magenta) and after (red) snap-through. Bottom: strain energy for circulator (blue) and reflector (red dashed). (E) Experimental circulator assembly into a mechanical topological metamaterial of two layers (I and II) connected via gemels; six co-rotational hinges (upper, middle, lower) prevent spatial interference.}
    \label{fig:1}
\end{figure}

Inspired by magnetically biased graphene, we propose bifurcation-based periodic lattices where angular momentum can propagate instabilities. This design enables controlled reflections and topological directional transport in a 2D array, opening new possibilities for elastic wave control. We begin with a straight, linear thin beam with perfect hinges at both ends (as depicted in Fig. \ref{fig:1}A). An axial force with slight moment is applied to induce its secondary buckling mode (S-shaped). When subjected to a downward load (blue arrow) at the beam midpoint, it snaps into a lower-energy configuration corresponding to its primary buckling mode (U-shaped). 
During this snap-through transition, asymmetric and distinct dynamics are witnessed on the two ends: the left (or right) end generates a significant (or passive) angular momentum with their maximum rotation angles $\alpha_r=39^{\circ}$ (or $\alpha_l=5^{\circ}$) corresponding to largest normalized input displacement ($u/L$, where $u$ is input displacement and $L$ is beam length). This contrast constitutes the origin of momentum biasing in our design.

Utilizing this principle, we design a three-port mechanical circulator comprising three elastic beams connected end-to-end via hinges to form an equilateral triangle, with beam midpoints serving as the ports. Compressing the circulator towards its centroid induces bifurcation into the secondary buckling mode, forming a chiral structure (Fig. \ref{fig:1}B). Applying a load at any port (e.g., Port 1) will trigger snap-through in that beam (correspondingly marked as beam 1). Due to momentum biasing, the deformation of hinges induces the adjacent beam 2 to snap into U-mode while beam 3 remains minimally deformed. Consequently, the output displacement at Port 2 significantly exceeds that of Port 3, demonstrating circulation. Crucially, the deformation of beam 1 and beam 2 generates a tension phenomenon in beam 3 and elevates its resistance to deformation. This negative feedback enhances the circulator's strong non-reciprocity (depicted in the middle panel of Fig. \ref{fig:1}D). Post-snap-through under full relaxation, the circulator's symmetry ensures transmission to the output port equals 1 and to the non-output port equals 0 at the minimum strain energy.

Second, inspired by acoustic topological insulators, we also construct a mechanical reflector. Adding a hard wall (see Fig. \ref{fig:1}C) outside a port (e.g., Port 2) blocks outward displacement propagation, thus inducing reflection and redirecting the signal to another port. Figure \ref{fig:1}D (top) plots port displacements versus normalized input displacement ($u_1/a$), where $u_1$ is the displacement on port 1 and $a$ is the side length of the circulator. Considering a circulator with the snapping position $|u_1/a|=0.063$, the displacement of Port 2 consistently exceeds Port 3 under negative load (towards the centroid), and this phenomenon is reversed under positive load (away from the centroid). For the reflector, the hard wall totally suppresses the outward motion of Port 2 and results in zero displacement. Under negative load (snapping position is $u_1/a=-0.13$), reflection causes Port 3 displacement to surpass Port 2 post-snap-through. For positive loads, where the signal bypasses Port 2, no reflection occurs and the response coincides with the circulator.

Figure \ref{fig:1}D (middle) shows transmission to Ports 2 and 3 versus normalized input displacement. Post-snap-through, a point pair indicates ideal circulator performance, which represents an absolutely releasing state ($|u_1/a|=0.12$). Remarkably, a similar pair exists pre-snap-through due to nonlinearity, representing a static circulator state recoverable upon load removal ($|u_1/a|=0.045$). Figure \ref{fig:1}D (bottom) depicts strain energy profiles. The circulator starts stable at $u_1/a=0$ ($E=650~\mathrm{J}$), undergoes minor energy increase (energy maximum: $|u_1/a|=0.046, E=660~\mathrm{J}$), snaps through (abrupt energy drop), reaches an energy minimum ($|u_1/a|=0.12, \,\, E=324~\mathrm{J}$), then rises sharply. The reflector requires substantially larger energy input before snap-through due to reflection (energy maximum: $u_1/a=-0.12, E=797~\mathrm{J}$). {The release of energy along with deformations enables time-reversal asymmetry of the system.}

Based on previous ingredients, we build a two-dimensional metamaterial. In Fig. \ref{fig:1}E, we assembled mechanical circulators into a mechanical topological structure comprising two layers. Adjacent circulators are connected at their ports via gemels. Additionally, as the rotation centers of all six circulators' hinges coincide, hinges within each layer are categorized into upper, middle, and lower types (See Figs. S1 to S5).

\begin{figure}
    \centering
    \includegraphics[width=\linewidth]{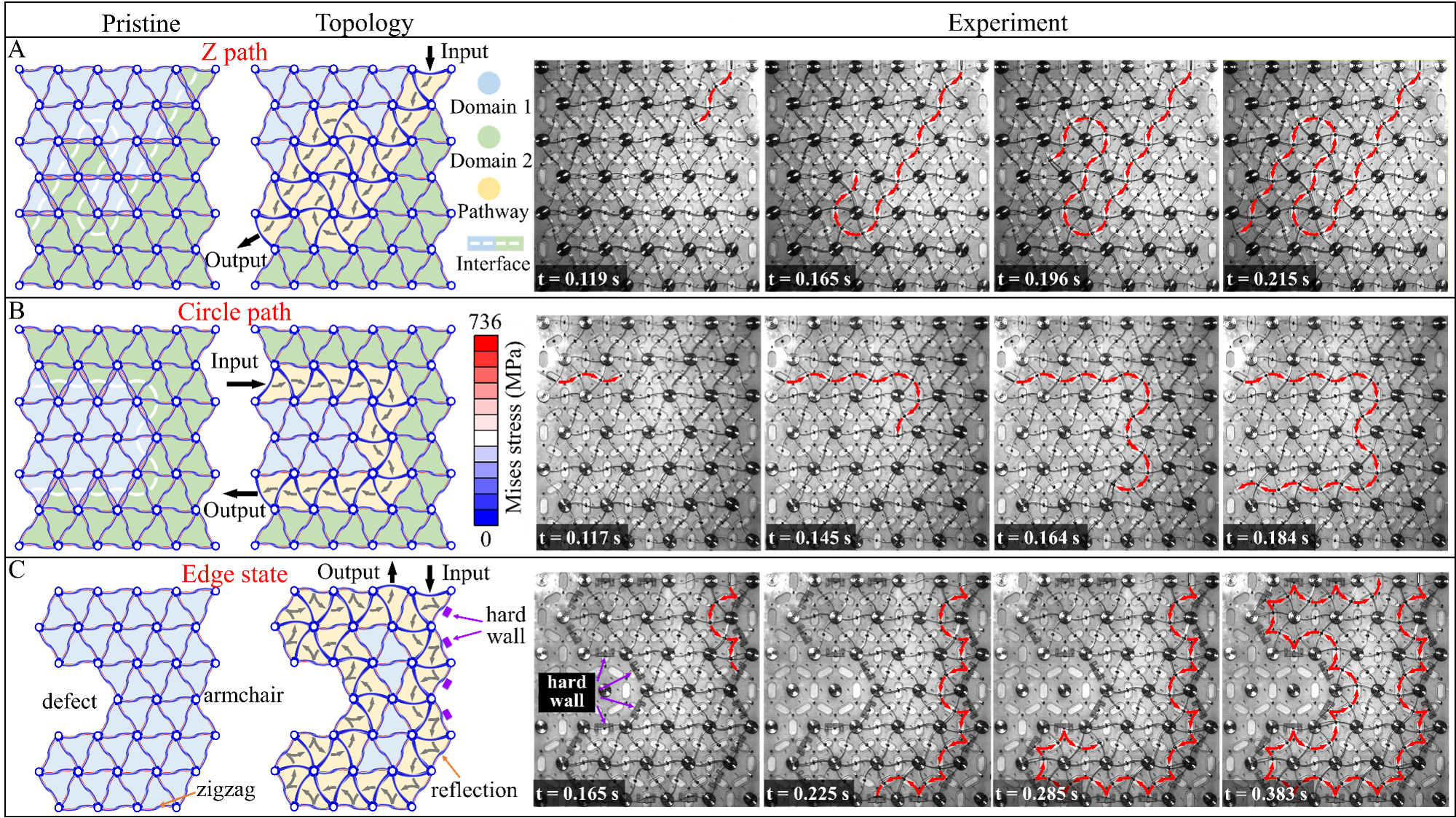}
    \caption{\textbf{Topological Insulator.} (A) Initial configuration of the mechanical topological domain wall Z path (left), deformed state (middle), and experimental snapshots at four time points (right group). Blue indicates clockwise circulators, green indicates counter-clockwise circulators, and yellow indicates deformed circulators. The white dashed line marks the interface between the two domains, and the grey arrow indicates the propagation direction. Stress distributions before and after deformation are shown in the left and middle images, respectively. Beams in the experiment were fabricated from thin spring steel. The parameters of the circulator are the same with Fig. 1. A total of 54 Circulators were used. (B) Mechanical topological domain wall circle path. Experimental parameters are identical to subfigure (A). (C) Mechanical topological edge state. A hard wall (purple block) was added at the medium boundary to reflect the displacement signal. Three boundary types were implemented: defect, armchair, and zigzag, to test the structure's defect immunity. A total of 41 circulators were used. All other parameters are identical to those stated above. Corresponding movies are presented in the supplementary videos S1 to S3.}
    \label{fig:2}
\end{figure}

Within the topological insulators depicted in Fig. \ref{fig:2}A, signals propagate unidirectionally along the interface between two distinct media without diffusing into the bulk body. Furthermore, this propagation occurs without backscattering due to defects or sharp bends along the path, demonstrating defect immunity. To validate the effectiveness of the mechanical topological domain wall, two complex paths were designed: the Z path (interface state A) and the circle path (interface state B), as shown in Fig. \ref{fig:2}A,B. Experimental images clearly show the displacement signal propagating stably along the media interface. The signal smoothly navigates sharp bends ($180^{\circ}$ for the Z path, $90^{\circ}$ for the circle path) without scattering. Additionally, the signal remains strictly confined to a width comparable to a single circulator during propagation, exhibiting no diffusion into the bulk medium. Leveraging these two key characteristics – defect immunity and tight confinement – the successful implementation of the Z path trajectory with a $180^{\circ}$ bend was achieved. In this trajectory, signals propagate sequentially between adjacent rows of circulators without mutual interference, a feat never seen in conventional wave systems. A mechanical topological edge state (interface state C) was also realized, as shown in Fig. \ref{fig:2}C. A hard wall (purple block), added at the boundary of the medium to reflect the displacement signal, was employed. Various boundary forms were implemented along this boundary: defect, armchair, and zigzag, to test the structure's defect immunity. Experimental images confirm the displacement signal successfully circumnavigating the boundary without being affected by the boundary form or defects. Crucially, the signal within the edge state is confined to a width of just one circulator. This extreme confinement allows the topological insulator to be highly compact. All experimental movies used to quantify the spatial and temporal properties are presented in the supplementary videos S1 to S3.

\begin{figure}
    \centering
    \includegraphics[width=\linewidth]{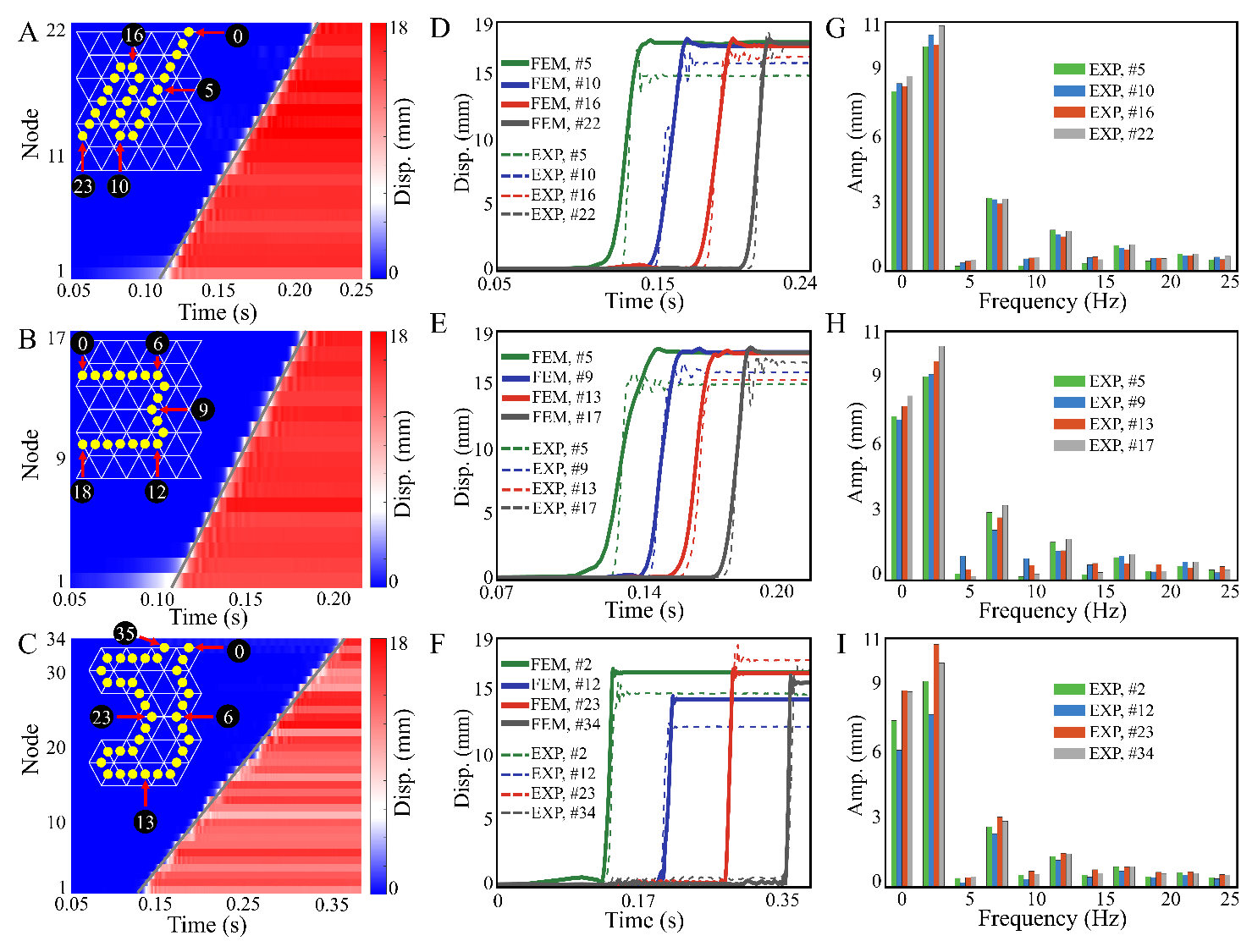}
    \caption{\textbf{Topological soliton propagation.} Experimental data for topological solitons propagating along the Z path (A), circle path (B), and the edge state (C). In panels A, B, and C, the horizontal axis represents time, the vertical axis represents the node coordinate index (excluding input and output nodes), with node numbering indicated in the insets. Blue indicates lower displacement, red indicates higher displacement. Propagation occurs at constant velocities: Z path $14.57~\mathrm{m/s}$, circle path $15.78~\mathrm{m/s}$, edge state $9.90~\mathrm{m/s}$. Grey lines indicate FEA simulation fits. The topological soliton waveforms for the Z path (D), circle path (E), and edge state (F) are also plotted. The horizontal axis is time, the vertical axis is soliton displacement. Dashed lines represent experimental results, solid lines represent FEA results. Soliton waveforms are shown for several nodes uniformly selected along the entire path. Stable soliton waveforms are maintained for the domain wall paths (Z path and circle path). For the edge state, the soliton waveforms show amplitude variation due to the pre-shifted hard wall, but no attenuation is observed. (G, H, I) To quantitatively verify soliton shape stability, Fourier transforms were performed on selected solitons to obtain their spectra. Comparison reveals highly consistent soliton spectra.}
    \label{fig:3}
\end{figure}

More quantitatively, a plunger was used to apply an impact at the input port with a speed of $90 ~\mathrm{mm/s}$. The propagation of the kink was captured using a high-speed camera, and the displacement at each circulator port (i.e., the location of the gemel) was measured using Digital Image Correlation (Fig. S7). Figure \ref{fig:3} shows experimental transition diagrams for the three cases. These diagrams represent the displacement over time at various nodes (i.e., circulator ports) along the path, demonstrating the kink propagating at a constant velocity. For all of the three interface states, their experimental displacements can be fitted as 
\begin{equation}
\centering
\begin{aligned}
u = A \left(( 1 - \frac{2}{\pi} \arctan \left[ e^{\frac{x - v(t-t_0)}{\sqrt{C_1 - {v^2}/{C_2}}}} \right] \right),
\label{kink1}
\end{aligned}
\end{equation}
where $A$ is the amplitude, $x$ is the position, $t$ is time, $t_0$ is the reference time, $v$ is the velocity, and $C_1$ and $C_2$ are constants. Such a displacement function proves to be a kink solution of the Sine-Gordon equation, which is given by  
\begin{equation}\label{kink2}
\centering
u_{tt} - u_{xx} + \sin(u) = 0,
\end{equation}
where $x$ denotes the spatial coordinate and $t$ denotes time. This further validates the nature of kink propagation in our topologically engineered lattices. More details are presented in Fig. S8.

For the domain wall paths (Z path and circle path), the displacements at different nodes were highly consistent, as shown in Fig. \ref{fig:3}A, B, D, and E. However, the edge state propagation process was more complex. Although Finite Element Analysis (FEA) simulations confirmed the feasibility of this process under ideal conditions, practical factors in the experiment, such as hinge friction and damping, hindered its propagation. Consequently, before the experiment, the hard wall was pre-shifted $7 ~\mathrm{mm}$ inwards towards the medium. This introduced significant initial internal stress to the circulators, making transitions easier for boundary circulators. However, it also resulted in a non-zero initial displacement at the path nodes, leading to non-uniform kink displacement amplitudes across nodes (Fig. \ref{fig:3}C, F). Despite this amplitude variation, the kink propagated over long distances without attenuation. To verify that the soliton maintained its shape during propagation, Fourier transforms were performed on solitons at specific positions. Fig. \ref{fig:3}G, H, I show that their spectra remained nearly identical, indicating stable soliton shape propagation. FEA simulations of the structure also showed excellent agreement with the experimental results.

\begin{figure}
    \centering
    \includegraphics[width=0.7\linewidth]{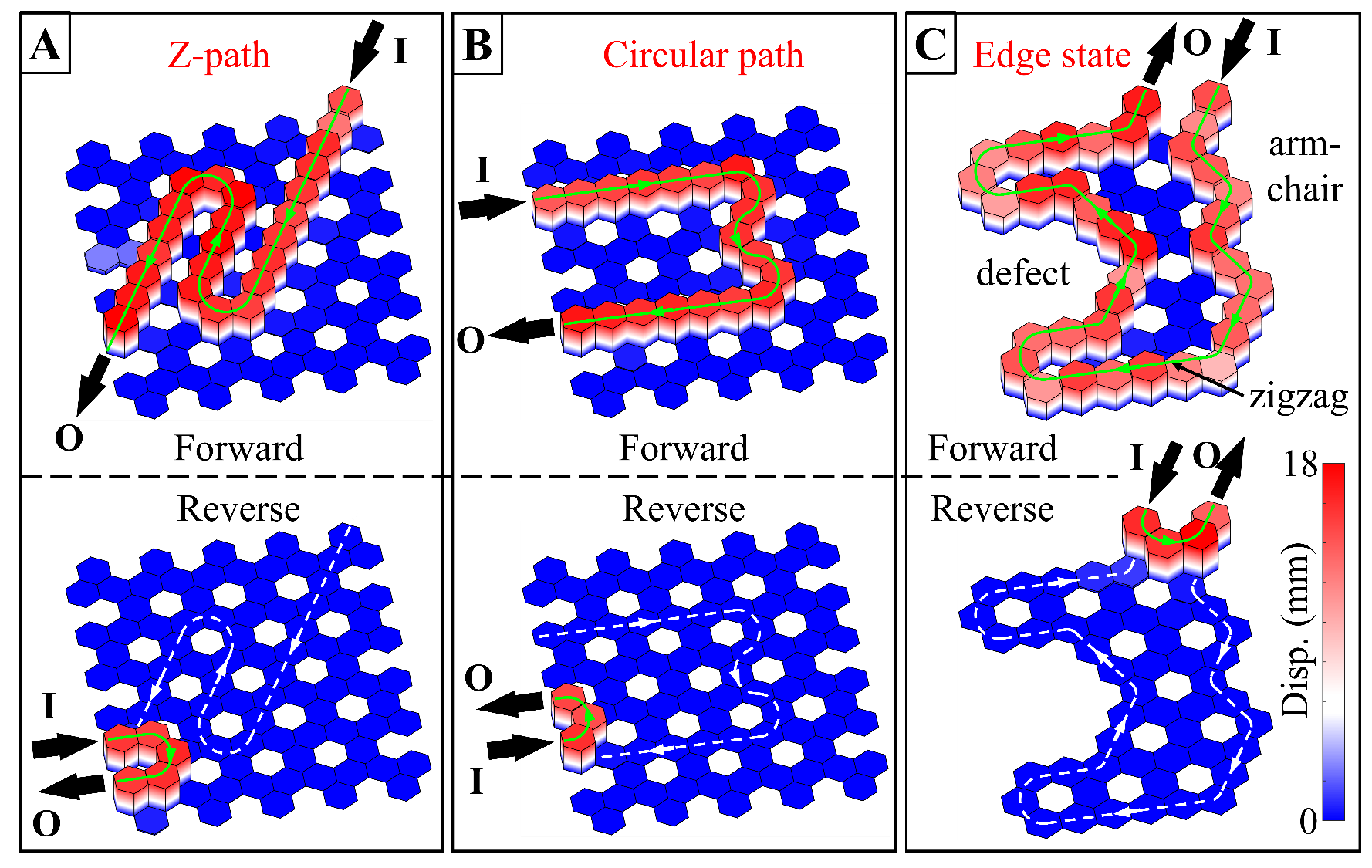}
    \caption{\textbf{Non-reciprocity of topological solitons}. Forward excitation (top row) and reverse excitation (bottom row) were applied to the domain wall Z path (A), domain wall circle path (B), and edge state (C), respectively. The corresponding displacement bar graphs are plotted. Distinct propagation pathways are observed for forward and reverse directions, confirming the non-reciprocity inherent to the mechanical spin-like Hall effect.}
    \label{fig:4}
\end{figure}

To verify the non-reciprocal propagation of the topological Hall insulator, a displacement excitation was subsequently applied at the output port location. The resulting propagation path under reverse excitation is shown in the bar graphs of Fig. \ref{fig:4}. The propagation path during reverse excitation was entirely distinct from that during forward propagation, confirming the unidirectional nature of the displacement signal.

Quantitatively, if we focus on the asymmetric rotational response of our mechanical circulator (see Fig. \ref{fig:1}B). The S-shaped beam from the first panel is incorporated into a triangular lattice with threefold rotational symmetry. When a force is applied at port 1, the snap-induced angular momentum at the 1–2 connection enables momentum transfer to port 2, noted by $\gamma_2$. Simultaneously, the lack of angular momentum between ports 1–3 results in low transfer to port 3, represented by $\gamma_1$. This forms the basis of non-ideal mechanical circulator. 
The scattering matrix of a non-ideal mechanical circulator is  
\begin{equation}
\centering 
\begin{bmatrix}
\begin{matrix} U^{\rm out}_1 \\ U^{\rm out}_2 \\ U^{\rm out}_3 \end{matrix} 
\end{bmatrix} =\begin{bmatrix}
\begin{matrix} 0 & \gamma_1 & \gamma_2 \\ \gamma_2 & 0 & \gamma_1 \\ \gamma_1 & \gamma_2 & 0 \end{matrix} 
\end{bmatrix}
\begin{bmatrix}
\begin{matrix} U^{\rm in}_1 \\ U^{\rm in}_2 \\ U^{\rm in}_3 \end{matrix} 
\end{bmatrix} 
\end{equation}

It immediately follows that the eigenvalues of the non-ideal circulator matrix are given by 
$\lambda_1 = \gamma_1 + \gamma_2$, 
$\lambda_2 = \gamma_1 e^{-i2\pi/3} + \gamma_2 e^{-i4\pi/3}$, 
and $\lambda_3 = \gamma_1 e^{-i4\pi/3} + \gamma_2 e^{-i2\pi/3}$. 
The stability condition, derived from energy minimisation and equilibrium considerations (see Fig. \ref{fig:1}D), drives the system toward its lowest-energy configuration once transient oscillations have dissipated. In this final steady state, the reverse transmission term $\gamma_2$ tends to zero, yielding the ideal circulator behaviour. In the ideal limit ($\gamma_1 = 1$, $\gamma_2 = 0$), the eigenvalue spectrum reduces to $\lambda = \{1,\, e^{-i2\pi/3},\, e^{-i4\pi/3}\}$, which corresponds to the perfect circularity of signal transmission among the three ports.

From this, it follows that two invariants can be clearly defined in our system: (i) the angle $\alpha$, which remains preserved, and (ii) the displacement of the midpoint of the buckling beam as it snaps from the second to the first buckling mode. The dynamics of this transition can be interpreted as a third, temporal invariant associated with the discrete, kink-like wave propagation.

Several critical aspects are essential in the proposed mechanical constructions. First, geometric and rotational symmetries are preserved to maintain topological robustness. The beam linkages enable self-rotation with minimal friction, while the supporting posts allow six-directional connections with reduced joint losses. All components remain within the linear elastic regime, avoiding plastic deformation, as verified by the normalized von Mises stress and strain (also see Fig. S6). 
Following experimental validation, kink-wave propagation was analyzed at different time intervals.

Time-dependent propagation of the generated kink perturbation reveals non-dispersive, conservative motion across all topological paths—interfaces A and B, and edge C—confirming the robustness of topological protection. Each selected unit cell exhibits pure, 
non-dispersive kink behavior, maintaining its shape and speed over time. 
Fourier analysis of node velocities further demonstrates discrete, stable eigenmodes, validating the quantized and non-dispersive nature of these kink waves.
Moreover, asymmetric propagation is observed: kink transmission is robust in one direction but nearly suppressed in the opposite, exhibiting strong nonreciprocity analogous to topological wave systems.

In summary, we present a new class of topologically protected mechanical metamaterials supporting nonreciprocal, kink-like wave propagation. By exploiting angular momentum and bifurcation-induced nonreciprocity, we realize mechanical circulators and reflectors 
that break time-reversal symmetry and achieve directionally controlled wave transport. 
Experiments and simulations confirm stable kink propagation and discrete frequency spectra, bridging nonlinear dynamics and topology in mechanical systems. 
This framework enables applications in vibration isolation, phononic routing, and energy harvesting, and paves the way for scalable, reconfigurable metamaterials for advanced sensing, energy transfer, and mechanical information storage.


\section*{Acknowledgements}
M.K. acknowledges support by the ANR PNanoBot [ANR-21-CE33-0015] and ANR OPTOBOTS project [ANR-21-CE33-0003]. Q.J. acknowledges support by Marie Skłodowska-Curie Actions Postdoctoral Fellowships [No. 101149710]. Data are available in the main text and the supplementary information.\\

\end{document}